\documentstyle[12pt]{article}

\newenvironment{eqar}[1]{\begin{equation}\label{#1}\begin{array}{rcl}}
{\end{array}\end{equation}}
\def\phan{\vphantom{\frac12}}
\def\phanc{\hphantom{(}}
\def\non{\nonumber}
\newcommand{\reff}[1]{(\ref{#1})}
\def\al{\alpha}
\def\be{\beta}
\def\ga{\gamma}
\def\la{\lambda}

\def\L{{\cal L}}

\author{S. M. Klishevich
\thanks{E-mail address: ASP9501@MX.IHEP.SU} \\
        {\it Institute for High Energy Physics} \\
        {\it Protvino, Moscow Region, 142284, Russia}}
\title{On redefinitions of variables in gauge field theory}
\date{June 1997}

\begin{document}
\maketitle

\begin{abstract}
 In this paper, for massive fields of spins 2 and 3 with
non-canonical Lagrangians, we build Hamiltonians and full systems of
constraints and show that the use of derivatives in a redefinition of
fields can give rise to a change of number of physical degrees of
freedom.
\end{abstract}

\newpage

\section*{Introduction}
At constructing various kind of field theory models it is often
useful to redefine initial fields for the theory to be of simpler and
more understandable form. Such substitutions of variables must not
change physical contents of the model i.e. the number of physical
degrees of freedom must remain the same as before the substitution.
In this, of course, it is meant that a modification of Poincare group 
representations didn't happen i.e. for example, a massive vector
field does not turn into three scalar fields. One has often to
perform such type of redefinitions in theories, which 
describe a physical particle with some set of fields, see
Ref.~\cite{Zinoviev-1,massive_spin}. So, for instance,
in~\cite{massive_spin} when describing massive spin-2 particle
propagation in a homogeneous electromagnetic field the result
independent of space-time dimensionality has been obtained using a
redefinition of second rank field.

One can divide all substitutions of variables into two kinds. First
kind are the substitutions of variables without derivatives i.e.
schematically $\Phi'_A=M_A^B\Phi_B+F_A^{BC}\Phi_B\Phi_C+\ldots$,
where $M_A^B$ is non-degenerate matrix. The second kind are
substitutions with derivatives i.e. they have form 
$\Phi'_A=M_A^B\Phi_B+H_A^B\partial\Phi_B+\ldots\ $. In this paper
using the case of free massive spin-3 field, we show that the number
of physical degrees of freedom of the theory can change, if one
uses derivatives in the redefinition of fields.

To begin with in Section 1 we consider the free massive spin-2 field
that is described with a non-canonical Lagrangian%
\footnote{We call a Lagrangian of free massive spin-$s$ field as
canonical, if it breaks into the sum of Lagrangians for massless
fields of spins $s,s-1,...,0$ in the massless limit,
see~\cite{massive_spin}.} {} derived from the canonical form with the
redefinition of the second rank field. We build a canonical
Hamiltonian and a full system of constraints%
\footnote{Describing systems with constraints, we use standard Dirac 
procedure, ref.~\cite{Dirac,Regge}.}
\ (all the constraints are of the first kind). A simple calculation
shows that the number of degrees of freedom remains the same 
at transition to the non-canonical form.

In Section 2 we are building a full system of constraints and
canonical Hamiltonian for a non-canonical Lagrangian, which describes
free massive spin-3 field and which is derived from the canonical
form with the redefinition of fields without derivatives. In this
case all the constraints are of the first kind and the number of
degrees of freedom remains the same as in the canonical case.

In Section 3 we consider the field of spin 3 in the non-canonical 
form that has been obtained from the canonical one with a
redefinition of fields with the use of the derivative. Building a
Hamiltonian and full systems of constraints, we show that in this
case the number of field degrees of freedom increases. In this the
constraints of the second kind are present in a full system.

\section{Free Massive Field with spin 2}
\label{spin2}
At first, let us consider the spin-2 field to compare with the case
of spin-3 field.

We consider the usual flat Minkowski space ${\bf M}^4$ with metric
signature $(1,-1,-1,-1)$. Latin indices take the value $k,l,... =
0,1,2,3$ and the Greek ones --- the value $\alpha,\beta,... = 1,2,3$.
For convenience we will not make difference between upper and lower
indices, while the summation over the repeated indices will be
understood, as usual, i.e.
$$
A_{k...}B_{k...}\equiv g^{kl}A_{k...}B_{l...} .
$$

We will describe the free massive field of spin 2 with the gauge
invariant Lagrangian of type
\begin{eqnarray}\label{l0}
{\cal L}_0 & = & \partial _m\bar h_{kl}\partial _mh_{kl}
\,-\,2\partial_kh_{kl}\partial _m\bar h_{lm}
\,+\,\left(\partial _kh_{kl}\partial _l\bar
h\,+\,h.c.\right)\,-\,\partial _k\bar h\partial _k h 
\non \\ &&{}
 +\,2\left(\partial _k\bar h_{kl}\partial _l\varphi 
\,-\,\partial _l\bar h\partial_l\varphi \,+\,h.c.\right)
\,-\,2\left(\partial _l\bar b_k\partial _lb_k\,-\,\partial
_lb_k\partial _k\bar b_l\right) \non \\  &&{}
+\,2m\left(\partial _l\bar b_kh_{kl}\,-\,\partial _k\bar b_kh
\,+\,h.c.\right)\,-\,m^2\left(\bar h_{kl}h_{kl}\,-\,\bar hh\right)\ ,
\end{eqnarray}
where $h_{kl}$ is a symmetrical tensor and $h=g^{kl}h_{kl}$.

The gauge transformations of the fields have the following form:
\begin{eqnarray}
\label{d0}
\delta h_{kl} & = & 2\partial _{(k}\xi _{l)}, \non \\
\delta b_k & = & \partial _k\eta \,+\,m\xi _k , \\ 
\delta \varphi & = & m\eta \non\ .
\end{eqnarray}

Lagrangian~\reff{l0} has been chosen in a non-canonical form (with
the off-diagonal kinetic part) in order that the Goldstone part
(proportional to mass) for the field $h_{kl}$ be absent in the
transformations. The transformations for $h$ has the form 
$\delta h_{kl} =2\partial_{(k} \xi_{l)} + m g_{kl}\eta$, where
$g_{kl}$ is the metrical tensor, in the canonical form with the same
normalization of fields. Therefore, to pass to Lagrangian~\reff{l0}
and transformations~\reff{d0} one need do the following substitution
of variables ${h'}_{kl} \to h_{kl}\,-\,g_{kl}\varphi$.

Further on, for convenience, we put $m=1$.

Passing from Lagrangian~\reff{l0} to the Hamiltonian formalism,
we get the following five constraints calculating the momenta
\begin{eqnarray}\label{cf2}
\stackrel{(1)}{C}{\!}^h_\alpha &=&p_{\alpha 0}^h
\,-\,\partial _\alpha h_{\beta \beta}
\,+\,2\partial _\beta h_{\alpha \beta }\,+\,2\partial _\alpha \varphi 
\,+\,\partial_\alpha h_{00}, \non \\
 \stackrel{(1)}{C}{\!}^h&=&p_{00}^h\,-\,\partial _\alpha h_{\alpha
0}, \\ 
\stackrel{(1)}{C}{\!}^b&=&p_0^b\,-\,2h_{\alpha \alpha } . \non\
\end{eqnarray}

Let us define the Poisson brackets in the following form:
\begin{eqnarray} 
\label{bra2}
\{h^{\al\be}\left(x\right),p^h_{\mu\nu}\left(y\right)\}&=&
\delta^{\phanc\al\be}_{(\mu\nu)}\left(x-y\right), \non \\
\{h^{\al0}\left(x\right),p^h_{\be0}\left(y\right)\}&=&
\delta^\al_\be\left(x-y\right), \non \\
\{h^{00}\left(x\right),p^h_{00}\left(y\right)\}&=&\delta\left(x-y
\right), \non \\
\{b^\al\left(x\right),p^b_{\be}\left(y\right)\}&=&
\delta^\al_\be\left(x-y\right), \non \\
\{b^0\left(x\right),p^b_0\left(y\right)\}&=&\delta\left(x-y\right),
\non \\
\{\varphi\left(x\right),p^\varphi\left(y\right)\}&=&\delta\left(x-y
\right)\ ,
\end{eqnarray}
where we use the notation $\delta^{\al\be...}_{\mu\nu...}\left(x-y
\right)\equiv\delta^\al_\mu\delta^{\be...}_{\nu...}\delta\left(x-y
\right)$.

The Poisson brackets of constraints~\reff{cf2} equal zero between 
themselves. At that the Hamiltonian obtained from~\reff{l0} has the 
following form:
\begin{eqnarray}\label{ham2}
{\cal H}_0&=&\bar p_{\alpha \beta }^hp_{\alpha \beta }^h-\frac 13\bar
p_{\beta\beta }^hp_{\alpha \alpha }^h+\frac 16\bar p_{\alpha \alpha
}^hp^\varphi +\frac 16p_{\alpha \alpha }^h\bar p^\varphi +\frac
12\bar p_\alpha^b p_\alpha ^b+\frac 16\bar p^\varphi p^\varphi  \non
\\
 &&{}+\frac 13 \partial _\alpha h_{\alpha 0}\bar p_{\beta \beta
}^h+\frac 13\partial _\alpha \bar h_{\alpha 0}p_{\beta \beta
}^h+\partial _\alpha \bar b_0p_\alpha ^b+\partial _\alpha
b_0\bar p_\alpha ^b-\frac 16\partial _\alpha h_{\alpha 0}\bar
p^\varphi  \non \\
 && {}-\frac 16\partial _\alpha \bar h_{\alpha 0}p^\varphi +\bar
h_{\alpha 0}p_\alpha ^b+h_{\alpha 0}\bar p_\alpha ^b-\partial _\alpha
h_{00}\partial _\beta \bar h_{\alpha \beta }-\partial _\beta \bar
h_{00}\partial _\alpha h_{\alpha \beta } \non \\
 && {}+\partial _\beta h_{\alpha \alpha }\partial _\beta \bar
h_{00}+\partial _\alpha h_{00}\partial _\alpha \bar h_{\beta \beta
}+\frac 23\partial _\alpha h_{\alpha 0}\partial _\beta \bar h_{\beta
0}-2\partial _\beta h_{\alpha 0}\partial _\beta \bar h_{\alpha 0}
\non\\ && \phan{}+\partial _\gamma h_{\alpha \beta }\partial _\gamma
\bar h_{\alpha \beta }+\partial _\beta h_{\alpha \alpha }\partial
_\gamma \bar h_{\beta \gamma }+\partial _\beta \bar h_{\gamma \gamma
}\partial _\alpha h_{\alpha \beta }-\partial _\beta h_{\alpha \alpha
}\partial _\beta \bar h_{\gamma \gamma }  \\
 && \phan {}-2\partial _\gamma \bar h_{\beta \gamma }\partial _\alpha
h_{\alpha \beta }-2\partial _\alpha \varphi \partial _\alpha \bar
h_{00}-2\partial _\alpha \bar \varphi \partial _\alpha
h_{00}+2\partial _\beta h_{\alpha \alpha }\partial _\beta \bar
\varphi  \non \\
 && \phan {}+2\partial _\beta \bar h_{\alpha \alpha }\partial _\beta
\varphi -2\partial _\beta \bar \varphi \partial _\alpha h_{\alpha
\beta }-2\partial _\alpha \bar h_{\alpha \beta }\partial _\beta
\varphi +2\partial _\beta b_\alpha \partial _\beta \bar b_\alpha
\non \\ && \phan {}-2\partial _\beta b_\alpha \partial _\alpha \bar
b_\beta -2\partial _\alpha b_\alpha \bar h_{00}-2\partial _\alpha
\bar b_\alpha h_{00}-2\partial _\beta b_\alpha \bar h_{\alpha \beta
}-2\partial _\beta \bar b_\alpha h_{\alpha \beta } \non \\
 && \phan {}+2\partial _\alpha b_\alpha \bar h_{\beta \beta
}+2\partial _\alpha \bar b_\alpha h_{\beta \beta }+4\partial _\alpha
\bar b_0h_{\alpha 0}+4\partial _\alpha b_0\bar h_{\alpha 0}+\bar
h_{\alpha \beta }h_{\alpha \beta }  \non \\
 &&  {}+\bar h_{00}h_{\alpha \alpha }+\bar h_{\alpha \alpha
}h_{00}-\bar h_{\beta \beta }h_{\alpha \alpha } \non\ ,
\end{eqnarray}
where $\ga_{\al\be}=-g_{\al\be}$ and 
$A_{\al\al...}\equiv \ga^{\al\be}A_{\al\be...}$.

In order that the Hamiltonian equations be equivalent to the
Lagrangian ones followed from~\reff{l0}, one has to add the first
step constraints to the Hamiltonian, but since the constraints
commute between themselves one needn't add it to Hamiltonian for the
calculation of second step constraints.

At the second stage we get 5 second step constraints, calculating
the evolution of first step ones
\begin{eqnarray}\label{cs2}
\stackrel{(2)}{C}{\!}^h_\alpha &=&2\partial _\beta p_{\alpha \beta
}^h \,-\,p_\alpha ^b\,-\,2\Delta h_{\alpha 0}\,-\,4\partial _\alpha
b_0 \non \\
\stackrel{(2)}{C}{\!}^h&=&{}-\,\Delta h_{\alpha \alpha }
\,+\,\partial _{\alpha \beta }h_{\alpha \beta}\,+\,2\Delta \varphi 
\,-\,2\partial _\alpha b_\alpha \,+\,h_{\alpha \alpha } \\ 
\stackrel{(2)}{C}{\!}^b&=&\partial _\alpha p_\alpha ^b\,-\,p^\varphi 
\,+\,2\partial _\alpha h_{\alpha 0} \non\ .
\end{eqnarray}

The Poisson brackets equal zero among the second step constraints and
between them and the first step ones.

The brackets between constraints~\reff{cs2} and
Hamiltonian~\reff{ham2} equal either zero or linear combinations of
second step constraints. That is, new constraints do not appear at
the third stage. Hence~\reff{cf2} and~\reff{cs2} form the full system
of constraints for this theory. In this, all the constraints are the
first kind ones.

It is easy to compute that the number of degrees of freedom equal
five in this case. This agrees with the formula $2s+1$ for a massive
particle of arbitrary spin $s$.

\section{Free massive field of spin 3: substitution of variables
without derivatives}
As in the previous Section we will describe a free massive complex
field of spin 3 with the gauge invariant Lagrangian in the
non-canonical form
\begin{eqnarray}
\label{l0s3a}
\L _0&=&{}-10\partial _n\bar \Phi _{klm}\partial _n\Phi
_{klm}+30\partial _k\Phi _{klm}\partial _n\bar \Phi _{lmn}-30\left(
\partial _k\Phi _{klm}\partial _m\bar \Phi _l+h.c.\right) \non \\
 &&  {}+30\partial _l\Phi _k\partial _l\bar \Phi _k+15\partial _l\bar
\Phi_l\partial _k\Phi _k-6\left(2\partial _k\Phi _{klm}\partial
_m\bar b_l- 2\partial_l\Phi _k\partial _l\bar b_k  \right. \non \\
 && \left. {}-\partial _l\bar \Phi _l\partial _kb_k+h.c.\right)+
\frac{36}5\partial _lb_k\partial _k\bar b_l+30\partial _m\bar
h_{kl}\partial _mh_{kl}-60\partial _kh_{kl}\partial _m\bar h_{lm} 
\non\\\non &&{} +30\left( \partial _l\bar h\partial _kh_{kl}+h.c.
\right)
-30\partial _kh\partial _k\bar h+5\left( \partial _l\bar \varphi
\partial _kh_{kl}-\partial _kh\partial _k\bar \varphi +h.c.\right)
\non \\
 && {}-\frac 14\partial _k\bar \varphi \partial _k\varphi -
15\left(2\partial _m \bar h_{kl}\Phi _{klm}-4\partial _k\bar
h_{kl}\Phi _l+\partial _k\bar h\Phi _k+h.c.\right) \non \\
 &&  {}-\frac 52\left( \partial _k\bar \varphi \Phi _k+h.c.\right)
-18\left( \partial _k\bar b_kh+h.c.\right) +5 \left(2\bar \Phi
_{klm}\Phi _{klm} \right. \non \\
 && \left. {}-6\bar \Phi_k\Phi _k +9\bar hh \right)\ ,
\end{eqnarray}
where $\Phi_{klm}$ is the symmetric tensor and
$\Phi_{k}\stackrel{def}{=}g^{lm} \Phi_{klm}$.

The transformations of the fields for this Lagrangian have the
following form:
\begin{eqar}{d0s3a}
\delta \Phi _{klm}&=&3\partial _{(k}\omega _{lm)}\,-\,
\frac35g_{(kl}\partial_{m)}\eta, \\ 
\delta h_{kl}&=&2\partial _{(k}\xi _{l)}\,+\,\omega _{kl}, \\
\delta b_k&=&2\partial _k\eta \,+\,5\xi _k, \\ 
\delta \varphi &=&12\eta ,
\end{eqar}
at that $g^{kl}\omega_{kl}=0$

The transformations of rank 2 and 3 fields have the form of type
$\delta \Phi =  \partial \omega + g\,\xi$ and $\delta h = \partial\xi
+ \omega + g\,\eta$. It is evident that transformations~\reff{d0s3a}
looks simpler, moreover the Goldstone part for the field $\Phi_{klm}$
is absent in the transformations. This facilitate an analysis of the
theory at switching on interaction. The transition from the canonical
form to Lagrangian~\reff{l0s3a} and transformations~\reff{d0s3a} has
been reached with the fields redefinitions of type
\begin{eqar}{subst3a}
\Phi'&\to&\Phi - g\,b, \\
 h'&\to& h - g\,\varphi\ .
\end{eqar}
In order to show that the number of degrees of freedom remains the
same we will compute the constraint algebra of theory~\reff{l0s3a}.

Calculating the canonically conjugated momenta we obtain 14 first
step constraints 
\begin{eqnarray} 
\label{cf3a}\non
\stackrel{(1)}{C}{\!\!}^b & = & \frac 65p_{000}^\Phi
-p_0^b-12\partial _\delta \Phi _{\gamma \gamma \delta }+24\partial
_\gamma \Phi _{\gamma 00}+ \frac{36}5\partial _\gamma b_\gamma , 
\\\phan 
\stackrel{(1)}{C}{\!\!}^h & = & {}-p_{00}^h-45\Phi _{\gamma \gamma
0}+15\Phi _{000}+30\partial _\gamma h_{\gamma 0}+18b_0, 
\\\phan \non
\stackrel{(1)}{C}{\!\!}_\alpha ^h & = &{}-p_{\alpha 0}^h+60\Phi
_{\alpha \gamma \gamma }-60\partial _\gamma h_{\alpha \gamma
}+30\partial _\alpha h_{\gamma \gamma }-30\partial _\alpha
h_{00}-5\partial _\alpha \varphi , 
\\\phan \non
\stackrel{(1)}{C}{\!\!}_\alpha ^\Phi & = &{}-p_{\alpha 00}^\Phi
+30\partial _\gamma \Phi _{\alpha \gamma 0}-30\partial _\alpha \Phi
_{\gamma \gamma 0}+30\partial _\alpha \Phi _{000}+12\partial _\alpha
b_0, 
\\\phan \non
\stackrel{(1)}{C}{\!\!}_{\alpha \beta }^\Phi  & = &{}-p_{\alpha \beta
0}^\Phi -3\ga_{\alpha \beta }p_{000}^\Phi -30\partial _\gamma \Phi
_{\alpha \beta \gamma }+30\partial _{(\alpha }\Phi _{\beta )\gamma
\gamma }-30\partial _{(\alpha }\Phi _{\beta )00} 
\\\phan  \non
&  &{}+30\ga_{\alpha \beta }\partial _\delta \Phi _{\gamma \gamma
\delta }-60\ga_{\alpha \beta }\partial _\gamma \Phi _{\gamma
00}-12\partial _{ (\alpha}b_{\beta )}-12\ga_{\alpha \beta }\partial
_\gamma b_\gamma \non\ .
\end{eqnarray}

Let us update Poisson brackets~\reff{bra2}
\begin{eqnarray}
\label{bra3a}
\{\Phi^{\al\be\ga}\left(x\right),p^\Phi_{\la\mu\nu}\left(y\right)\}
&=& \delta^{\phanc\al\be\ga}_{(\la\mu\nu)}\left(x-y\right),
\non\\
\{\Phi^{\al\be0}\left(x\right),p^\Phi_{\mu\nu0}\left(y\right)\}&=&
\delta^{\phanc\al\be}_{(\mu\nu)}\left(x-y\right),
\non\\
\{\Phi^{\al00}\left(x\right),p^\Phi_{\be00}\left(y\right)\}&=&
\delta^\al_\be\left(x-y\right),
\non\\
\{\Phi^{000}\left(x\right),p^\Phi_{000}\left(y\right)\}&=&
\delta\left(x-y\right)\ .
\end{eqnarray}
The Poisson brackets of all the first step constraints equal zero
among themselves.

Now we need to compute the canonical Hamiltonian. The result is
rather cumbersome even for free field, therefore, we place the
concrete expression for the Hamiltonian in Appendix~\ref{ap-simple}.

From the condition of conservation of the first step constraints, we
get the second step constraints
\begin{eqnarray} 
\label{cs3a}\non
\stackrel{(2)}{C}{\!\!}^b & = &{}-\partial _\gamma p_\gamma
^b-0.1p_{\gamma \gamma }^h-5.4b_0+6p^\varphi -24\partial ^2\Phi
_{\gamma \gamma 0}+12\partial _{\gamma \delta }^2\Phi _{\gamma \delta
0}
\\\phan \non && 
{}+12\partial ^2\Phi_{000}+7.2\partial ^2b_0-33\partial _\gamma
h_{\gamma 0} +13.5\Phi _{\gamma\gamma 0}-4.5\Phi _{000}, 
\\\phan \non 
\stackrel{(2)}{C}{\!\!}^h & = & 3p_{000}^\Phi -30\partial ^2h_{\gamma
\gamma }+30\partial _{\gamma \delta }^2h_{\gamma \delta }+5\partial
^2\varphi -60\partial _\delta \Phi _{\gamma \gamma \delta
}+90\partial _\gamma \Phi _{\gamma 00}
 \\\phan \non &&
{}+45h_{\gamma \gamma }-45h_{00}, 
\\\phan \non 
\stackrel{(2)}{C}{\!\!}_\alpha ^h & = &{}-2\partial _\gamma p_{\alpha
\gamma }^h+5p_\alpha ^b+60\partial ^2h_{\alpha 0}-60\partial _\gamma
\Phi _{\alpha \gamma 0}+30\partial _\alpha \Phi _{\gamma \gamma 0} 
\\\phan &&
{}-30\partial _\alpha \Phi_{000}, 
\\\phan \non 
\stackrel{(2)}{C}{\!\!}_\alpha ^\Phi & = & 3\partial _\alpha
p_{000}^\Phi -30\partial ^2\Phi _{\alpha \gamma \gamma }+30\partial
_{\gamma \delta }^2\Phi _{\alpha \gamma \delta }-60\partial _{\alpha
\delta }^2\Phi _{\gamma \gamma \delta }+90\partial _{\alpha \gamma
}^2\Phi _{\gamma 00} 
\\\phan \non &&
{}+12\partial^2b_\alpha +24\partial _{\alpha \gamma }^2b_\gamma 
+30\Phi _{\alpha \gamma \gamma} -60\partial _\gamma h_{\alpha \gamma
}+15\partial _\alpha h_{\gamma \gamma}
\\\phan \non &&
{}-45\partial _\alpha h_{00}-\frac 52\partial _\alpha \varphi , 
\\\phan \non 
\stackrel{(2)}{C}{\!\!}_{\alpha \beta }^\Phi & = &{}-3\partial
_\gamma p_{\alpha \beta \gamma }^\Phi +p_{\alpha \beta }^h+30\partial
^2\Phi _{\alpha \beta 0}-30\partial _{\alpha \beta }^2\Phi
_{000}+30\partial _{\alpha \beta }^2\Phi _{\gamma \gamma 0} 
\\\phan \non &&
{}-30\ga_{\alpha \beta }\partial_{\gamma \delta }^2\Phi _{\gamma
\delta 0} -30\ga_{\alpha \beta }\partial ^2\Phi _{000}+60\ga_{\alpha
\beta }\partial ^2\Phi _{\gamma \gamma 0}-12\partial _{\alpha \beta
}^2b_0 
\\\phan \non &&
{}-24\ga_{\alpha \beta}\partial ^2b_0+60\partial _{(\alpha }h_{\beta
)0} +90\ga_{\alpha \beta }\partial _\gamma h_{\gamma 0}-45\ga_{\alpha
\beta }\Phi _{\gamma \gamma 0} 
\\\phan \non &&
{}+15\ga_{\alpha \beta }\Phi _{000}+18\ga_{\alpha \beta}b_0\ .
\end{eqnarray}

The second step constraints have zero brackets among themselves and
between them and the first step ones. New constraints do not appear
at the third stage. Hence \reff{cf3a} and~\reff{cs3a} constitute the
full system of constraints. In this, all the constraints are of the
first kind.

It is easy to compute the number of independent field degrees of
freedom. The number of all field components equals 35 and number of
the constraints 28, therefore, the number of independent degrees of
freedom equals $35-28=7$. Thus passing to the non-canonical
form~\reff{l0s3a} with the substitution of variables~\reff{subst3a},
the number of degrees of freedom has not changed.

\section{Massive spin-3 field: substitution with de\-ri\-vatives}

When looking at transformations~\reff{d0s3a} a desire arises to
simplify the ones making a third rank field shift of type
\begin{equation}
\label{shift3b}
\Phi'\to\Phi + g\partial\varphi
\end{equation}
so that the transformations for $\Phi$ remain only of type
$\partial\omega$. Besides, simplicity of the transformations gives us
another advantage. Since the metrical tensor is absent in the
transformations after such shift, the Lagrangian does not depend on
the space-time dimensionality.

However the Lagrangian becomes the third degree one in derivatives.
The question emerges whether the number of physical degrees of
freedom changes at that.

 Let us show that the number of degrees of freedom increases by one
at the redefinitions of type~\reff{shift3b}.

In order to reduce the number of derivatives in the Lagrangian we
introduce an auxiliary field $v_k$. In this, the Lagrangian acquire
the following form
\begin{eqnarray}
\label{l0s3b} \non
\L_0&=&\left( 2\partial _m\Phi _{klm}\partial _l\bar v_k-2\partial
_l\Phi _k\partial _l\bar v_k+2\partial _l\Phi _k\partial _k\bar
v_l-3\partial _l\Phi _l\partial _m\bar v_m+h.c.\right)  
\\ && \non
{}-10\partial _n\Phi _{klm}\partial _n\bar \Phi _{klm}+30\partial
_n\Phi _{kln}\partial _m\bar \Phi _{klm}-30\left( \partial _n\Phi
_{kmn}\partial _m\bar \Phi _k+h.c.\right)  
\\ && \non 
{}+30\partial _m\Phi _k\partial _m\bar \Phi _k+15\partial _m\Phi
_m\partial _k\bar \Phi _k-6\left(2\partial _m\Phi _{klm}\partial
_l\bar b_k-2\partial _mb_l\partial _m\bar \Phi _l  \right.
 \\ &&\non \left.
\!{}-\partial _mb_m\partial _l\bar \Phi _l+h.c.\right)+30\partial
_mh_{kl} \partial_m\bar h_{kl}-60\partial _mh_{km}\partial _l\bar
h_{kl} 
\\ && \non
{}+30\left(\partial_mh_{lm}\partial _l\bar h 
+h.c.\right)-30\partial _lh\partial _l\bar h-15\left( 2\partial
_mh_{kl} \bar\Phi_{klm} \right. 
\\ && \non \left.
\!{}-4\partial _mh_{km}\bar \Phi _k+\partial _kh\bar \Phi
_k+h.c.\right) +\left( \bar \lambda _k\left( \partial _k\varphi
-v_k\right) +h.c.\right)
\\ &&
{}+10\Phi _{klm}\bar \Phi _{klm}-30\Phi _k\bar \Phi _k\ .
\end{eqnarray}
Correspondingly, gauge transformations~\reff{d0s3a} after 
shift~\reff{shift3b} and entering the auxiliary field have the
following form:
\begin{eqar}{d0s3b}
\delta\Phi_{klm}&=&3\partial_{(k}\omega_{lm)},\\
\delta h_{kl}&=&2\partial_{(k}\xi_{l)}+\omega_{kl},\\
\delta b_{k}&=&2\partial_k\eta+5\xi_k,\\
\delta v_{k}&=&12\partial_k\eta,\\
\delta\varphi&=&12\eta\ .
\end{eqar}

Passing to the Hamiltonian form of theory~\reff{l0s3b}, we obtain the 
following constraints at this stage
\begin{eqnarray}
\label{cf3b}
\stackrel{(1)}{C}{\!\!}^\lambda & = & p_0^\lambda , \non\\ 
\stackrel{(1)}{C}{\!\!}_\alpha ^\lambda & = & p_\alpha ^\lambda ,
\non\\ 
\stackrel{(1)}{C}{\!\!}^\varphi & = & p^\varphi \,-\,\lambda _0\,,
\non\\ 
\stackrel{(1)}{C}{\!\!}^v & = &{}-\frac 16
\,p_0^b\,-\,p_0^v\,-\,2\partial
_\delta \Phi _{\gamma \gamma \delta }\,+\,2\partial _\gamma \Phi
_{\gamma00}, 
\non\\ 
\stackrel{(1)}{C}{\!\!}_\alpha ^v & = &{}-\frac 16\,p_\alpha
^b\,-\,p_\alpha ^v\,+\,2\partial _\alpha \Phi _{\gamma \gamma
0}\,-\,2\partial _\alpha \Phi _{000}, 
\non\\ 
\stackrel{(1)}{C}{\!\!}^h & = &{}-\,p_{00}^h\,-\,45\Phi _{\gamma
\gamma 0}\,+\,15\Phi _{000}\,+\,30\partial _\gamma h_{\gamma 0}, 
\non\\ 
\stackrel{(1)}{C}{\!\!}_\alpha ^h & = &{}-\,p_{\alpha 0}^h\,+\,60\Phi
_{\alpha \gamma \gamma }\,-\,60\partial _\gamma h_{\alpha \gamma
}\,+\,30\partial _\alpha h_{\gamma \gamma }\,-\,30\partial _\alpha
h_{00},
\non\\ 
\stackrel{(1)}{C}{\!\!}_\alpha ^\Phi & = &{}-\,p_{\alpha 00}^\Phi
\,+\,30\partial _\gamma \Phi _{\alpha \gamma 0}\,-\,30\partial
_\alpha \Phi _{\gamma \gamma 0}, \non\\
\stackrel{(1)}{C}{\!\!}_{\alpha \beta }^\Phi  & = &{}-\,p_{\alpha
\beta 0}^\Phi \,-\,3p_{000}^\Phi \ga_{\alpha \beta }\,-\,30\partial
_\gamma \Phi _{\alpha \beta \gamma }\,-\,30\partial _{(\alpha }\Phi
_{\beta )00}\,+\,30\partial _{(\alpha }\Phi _{\beta )\gamma \gamma } 
\non\\ &&
{}+\,30\partial_\delta \Phi _{\gamma \gamma \delta }\ga_{\alpha \beta
}  \,-\,60\partial _\gamma \Phi _{\gamma 00}\ga_{\alpha \beta}
\,-\,12\partial _{(\alpha }b_{\beta )}\,-\,12\partial _\gamma
b_\gamma \ga_{\alpha \beta } 
\non\\ &&
{}+\,2\partial _{(\alpha }v_{\beta )}\,+\,6\partial _\gamma
v_\gamma \ga_{\alpha \beta }\ .
\end{eqnarray}

Since unlike~\reff{l0s3a} the additional variables, namely, the
auxiliary field $v_k$ and the Lagrange multiplier $\la_k$ arise in
Lagrangian~\reff{l0s3b}, therefore, one has to update the Poisson
brackets
\begin{eqnarray}
\label{bra3b}
\{v_0\left(x\right),p^v_0\left(y\right)\}&=&\delta\left(x-y\right),
\non\\
\{v_\al\left(x\right),p^v_\be\left(y\right)\}&=&
\delta_{\al\be}\left(x-y\right),
\non\\
\{\la_0\left(x\right),p^\la_0\left(y\right)\}&=&\delta\left(x-y\right
),
\non\\
\{\la_\al\left(x\right),p^\la_\be\left(y\right)\}&=&
\delta_{\al\be}\left(x-y\right)\ .
\end{eqnarray}

There are only two the non-trivial brackets among the first step
constraints
\begin{equation}
\label{alg3b1}
\{\stackrel{(1)}{C}{\!\!}^\la\left(x\right),
\stackrel{(1)}{C}{\!\!}^v\left(y\right)\}=\delta\left(x-y\right)\ ,
\end{equation}
hence, besides first kind constraints, the second kind ones emerge in
the theory.

Canonical Hamiltonian obtained in this case is placed in
appendix~\ref{ap-ham}.

From the condition of the first step constraint conservation, we
obtain the second step constraints
\begin{eqnarray}
\label{cs3b}
\stackrel{(1)}{\Lambda}{\!\!}_0^\lambda & = & \partial _\alpha
\lambda _\alpha , \non\\ 
\stackrel{(1)}{\Lambda}{\!\!}^\varphi & = & v_0, \non\\ 
\stackrel{(2)}{C}{\!\!}_\alpha ^\lambda & = & \,-\,\partial _\alpha
\varphi \,+\,v_\alpha , \non\\ 
\stackrel{(2)}{C}{\!\!}^v & = & \lambda _0, \non\\ 
\stackrel{(2)}{C}{\!\!}_\alpha ^v & = & \lambda _\alpha , \non\\ 
\stackrel{(2)}{C}{\!\!}^h & = & \frac 52p_0^b\,-\,30\partial
^2h_{\gamma \gamma }\,+\,30\partial _{\gamma \delta }^2h_{\gamma
\delta }\,-\,30\partial _\delta \Phi _{\gamma \gamma \delta
}\,+\,30\partial _\gamma \Phi _{\gamma 00},
 \non\\ 
\stackrel{(2)}{C}{\!\!}_\alpha ^h & = &{}-\,2\partial _\gamma
p_{\alpha \gamma }^h\,+\,5p_\alpha ^b\,+\,60\partial ^2h_{\alpha
0}\,-\,60\partial _\gamma \Phi _{\alpha \gamma 0}\,+\,30\partial
_\alpha \Phi _{\gamma \gamma0}
\non\\ &&
{}-\,30\partial _\alpha \Phi _{000}, \non\\ 
\stackrel{(2)}{C}{\!\!}_\alpha ^\Phi  & = & {}+\,3\partial _\alpha
p_{000}^\Phi \,-\,30\partial ^2\Phi _{\alpha \gamma \gamma
}\,+\,30\partial _{\gamma \delta }^2\Phi _{\alpha \gamma \delta
}\,-\,60\partial _{\alpha \delta }^2\Phi _{\gamma \gamma \delta }
\,+\,90\partial _{\alpha \gamma}^2\Phi _{\gamma 00} 
\non\\ &&
{}+\,12\partial ^2b_\alpha  \,+\,24\partial _{\alpha \gamma
}^2b_\gamma \,-\,2\partial ^2v_\alpha \,-\,10\partial _{\alpha \gamma
}^2v_\gamma \,+\,30\Phi _{\alpha \gamma \gamma } 
\non\\ &&
{}-\,60\partial _\gamma h_{\alpha \gamma } \,+\,15\partial _\alpha
h_{\gamma \gamma }\,-\,45\partial _\alpha h_{00},
 \non\\ 
\stackrel{(2)}{C}{\!\!}_{\alpha \beta }^\Phi & = &{}-\,3\partial
_\gamma p_{\alpha \beta \gamma }^\Phi \,+\,p_{\alpha \beta
}^h\,+\,30\partial ^2\Phi _{\alpha \beta 0}\,-\,30\partial _{\alpha
\beta }^2\Phi _{000}\,+\,30\partial _{\alpha \beta }^2\Phi _{\gamma
\gamma0} 
\non\\ &&
{}+\,60\ga_{\alpha \beta }\partial ^2\Phi _{\gamma \gamma 0}
 \,-\,30\ga_{\alpha \beta }\partial ^2\Phi _{000}\,-\,30\ga_{\alpha
\beta }\partial _{\gamma \delta }^2\Phi _{\gamma \delta
0}\,-\,12\partial _{\alpha \beta }^2b_0 
\non\\ &&
{}-\,24\ga_{\alpha \beta }\partial ^2b_0\,+\,6\partial _{\alpha\beta
}^2v_0  \,+\,6\ga_{\alpha \beta }\partial ^2v_0\,+\,60\partial
_{(\alpha} h_{\beta )0}
 \non\\ &&
{}+\,90\ga_{\alpha \beta }\partial _\gamma h_{\gamma0}
\,+\,15\ga_{\alpha \beta }\Phi _{000}\,-\,45\ga_{\alpha \beta }\Phi
_{\gamma \gamma 0}\ .
\end{eqnarray}

The first and second step constraints besides~\reff{alg3b1} have the 
following non-trivial Poisson brackets
\begin{eqnarray}
\label{alg3b2}
\{\stackrel{(1)}{C}{\!\!}^\la\left(x\right),
\stackrel{(2)}{C}{\!\!}^v\left(y\right)\}&=&\delta\left(x-y\right),
\non\\
\{\stackrel{(1)}{C}{\!\!}^\varphi\left(x\right)
\stackrel{(2)}{C}{\!\!}^\la_\al\left(y\right)\}&=&
\partial^y_\al\delta\left(x-y\right),
\non\\
\{\stackrel{(1)}{C}{\!\!}^v_\al\left(x\right)
\stackrel{(2)}{C}{\!\!}^\la_\be\left(y\right)\}&=&\delta_{\al\be}
\left(x-y \right),
\non\\
\{\stackrel{(1)}{C}{\!\!}^\la_\al\left(x\right)
\stackrel{(2)}{C}{\!\!}^v_\be\left(y\right)\}&=&{}
-\delta_{\al\be}\left(x-y\right)\ .
\end{eqnarray}
where $\partial^y_\al=\frac{\partial}{\partial y_\al}$.

At the third stage new constraints do not emerge but the partial
determination of the Lagrange multipliers happens
$$
\Lambda^v_\al={}-\partial_\al v_0,\ \Lambda^\la_\al=0 .
$$

Thus, we have 15 "non-commutative" constraints. These are the
first step constraints$
\stackrel{(1)}{C}{\!\!}^\varphi,
\stackrel{(1)}{C}{\!\!}^\al,
\stackrel{(1)}{C}{\!\!}^\la_\al,
\stackrel{(1)}{C}{\!\!}^v_\al
$
and the second step ones
$
\stackrel{(2)}{C}{\!\!}^\la_\al,
\stackrel{(2)}{C}{\!\!}^\varphi
\stackrel{(2)}{C}{\!\!}^\varphi_\al
$.
Among these constraints there is a linear combination, that has zero
brackets with all other constraints, i.e., it is the first kind
constraint
$$
\stackrel{(1)}{C}{\!\!}^c\,+\,\partial_\alpha\!\!
\stackrel{(1)}{C}{\!\!}^v_\alpha\,+\,
\stackrel{(2)}{C}{\!\!}^v ,
$$
thus, among 15 "non-commutative" constraints, there are only
14 second kind ones.

Having computed the constraint algebra, let us calculate the number of
degrees of freedom. The number of all field components in
theory~\reff{l0s3b} equals $20+10+4+1+4+4=43$. In this, there are 22
first and 20 second step constraints. Among them, there are 28 first
and 14 second kind constraints. Hence the number of degrees of
freedom for this theory equals $43-28-{14\over2}=8$ and not 7 as for
the theory describing the massive particle of spin 3.

Thus, one can conclude that the presence of derivatives in field
redefinitions, as in~\reff{shift3b} for example, can result in the
change of the number of degrees of  freedom in the theory.

\section{Conclusion}

Thus, in this paper we built the canonical Hamiltonians and full
systems of constraints for the free massive fields of spin 2 and 3.
We have showen that at substitutions of variables with use of
derivatives the number of physical degrees of freedom in theory will
be able to change. Of course it is not mean that such changes always
happen. It implies that the use derivatives in substitutions of
variables requires more careful examination.

The author would like to thank prof.~Yu.~M.~Zinoviev for useful
discussion and help in the work. Work supported by Russian Foundation
for Fundamental Research grant 95-02-06312.


\appendix
\section{}
\label{ap-simple}

The canonical Hamiltonian for the Lagrangian~\reff{l0s3a} has the
following form:
\begin{eqnarray}\non
{\cal H}&=&\frac 15\bar p_{000}^\Phi p_{000}^\Phi +\frac 1{10}\bar
p_{\alpha \beta \gamma }^\Phi p_{\alpha \beta \gamma }^\Phi -\frac
3{50}\bar p_{\beta \gamma \gamma }^\Phi p_{\alpha \alpha \beta }^\Phi
+\frac 1{20}\bar p_{\alpha \alpha \beta }^\Phi p_\beta ^b+\frac
1{20}\bar p_\beta ^bp_{\alpha \alpha \beta }^\Phi 
 \\\phan \non &&{} +\frac 1{30}\bar p_{\alpha \beta }^h
p_{\alpha \beta }^h{}-\frac 7{720}\bar p_{\beta \beta }^hp_{\alpha
\alpha }^h+\frac 1{12}\bar p_{\alpha \alpha }^hp^\varphi +\frac
1{12}p_{\alpha \alpha }^h\bar p^\varphi +\frac 16\bar p_\alpha
^bp_\alpha ^b+\bar p^\varphi p^\varphi  \\
\phan \non && 
{}-3\partial _\beta \Phi _{\alpha \alpha \beta }\bar p_{000}^\Phi {}
-3\partial _\beta \bar \Phi _{\alpha \alpha \beta }p_{000}^\Phi
+3\partial _\alpha \Phi _{\alpha 00}\bar p_{000}^\Phi +3\partial
_\alpha \bar \Phi _{\alpha 00}p_{000}^\Phi
  \\\phan  && 
{}+\frac 35\partial _\alpha \Phi _{\alpha \beta 0}\bar p_{\beta
\gamma \gamma }^\Phi +\frac 35\partial _\alpha \bar \Phi _{\alpha
\beta 0}p_{\beta \gamma \gamma }^\Phi {}+\frac 65\partial _\alpha
b_\alpha \bar p_{000}^\Phi +\frac 65\partial _\alpha \bar b_\alpha
p_{000}^\Phi  \\\phan \non && {}+\frac 9{25}\partial _\alpha b_0\bar
p_{\alpha \beta \beta }^\Phi +\frac 9{25}\partial _\alpha \bar
b_0p_{\alpha \beta \beta }^\Phi +\frac 7{24}\partial _\alpha
h_{\alpha 0}\bar p_{\beta \beta }^h{}+\frac 7{24}\partial _\alpha
\bar h_{\alpha 0}p_{\beta \beta }^h
 \\\phan \non && 
{}-\frac 12\partial _\alpha \Phi _{\alpha \beta 0}\bar p_\beta
^b-\frac 12\partial _\alpha \bar \Phi _{\alpha \beta 0}p_\beta
^b+\frac 65\partial _\alpha b_0\bar p_\alpha ^b+\frac 65\partial
_\alpha \bar b_0p_\alpha ^b-\frac 52\partial _\alpha h_{\alpha 0}\bar
p^\varphi
  \\\phan \non && 
{}-\frac 52\partial _\alpha \bar h_{\alpha 0}p^\varphi +\frac
7{48}\bar \Phi _{000}p_{\alpha \alpha }^h+\frac 7{48}\Phi _{000}\bar
p_{\alpha \alpha }^h+\bar \Phi _{\alpha \beta 0}p_{\alpha \beta
}^h+\Phi _{\alpha \beta 0}\bar p_{\alpha \beta }^h
 \\\phan \non && 
{}-\frac 7{16}\bar \Phi _{\alpha \alpha 0}p_{\beta \beta }^h{}-\frac
7{16} \Phi_{\alpha \alpha 0}\bar p_{\beta \beta }^h+\frac 3{40}\bar
b_0p_{\alpha \alpha }^h+\frac 3{40}b_0\bar p_{\alpha \alpha }^h-\frac
54\bar \Phi _{000}p^\varphi 
 \\\phan \non && 
{}-\frac 54\Phi _{000}\bar p^\varphi + \frac{15}4\bar \Phi _{\alpha
\alpha 0}p^\varphi +\frac{15}4\Phi _{\alpha \alpha 0}\bar p^\varphi
{}+\frac 92\bar b_0p^\varphi +\frac 92b_0\bar p^\varphi  \\\phan \non
&&{}+30\partial _\beta \Phi _{\alpha \alpha \beta } \partial
_\delta \bar \Phi _{\gamma \gamma \delta }-30\partial _\beta \Phi
_{\alpha \alpha \beta }\partial _\gamma \bar \Phi _{\gamma
00}-12\partial _\beta \Phi _{\alpha \alpha \beta }\partial _\gamma
\bar b_\gamma
  \\\phan \non && 
{}+10\partial _\delta \Phi _{\alpha \beta \gamma }\partial _\delta
\bar \Phi _{\alpha \beta \gamma }+30\partial _\gamma \Phi _{\alpha
\alpha \beta }\partial _\delta \bar \Phi _{\beta \gamma \delta
}-30\partial _\gamma \Phi _{\alpha \alpha \beta }\partial _\gamma
\bar \Phi _{\beta \delta \delta } 
\\\phan\non && 
{}+30\partial _\gamma \Phi _{\alpha \alpha \beta }\partial _\gamma
\bar \Phi _{\beta 00}{}+12\partial _\gamma \Phi _{\alpha \alpha \beta
}\partial _\gamma \bar b_\beta -30\partial _\delta \bar \Phi _{\beta
\gamma \delta }\partial _\alpha \Phi _{\alpha \beta \gamma }
 \\\phan \non &&  
{}+30\partial _\gamma \bar \Phi _{\beta \delta \delta }\partial
_\alpha \Phi _{\alpha \beta \gamma }-30\partial _\gamma \bar \Phi
_{\beta 00}\partial _\alpha \Phi _{\alpha \beta \gamma }{}-12\partial
_\alpha \Phi _{\alpha \beta \gamma }\partial _\gamma \bar b_\beta
\\\phan \non &&  
{}-12\partial _\beta \bar \Phi _{\alpha \alpha \beta }\partial
_\gamma b_\gamma +12\partial _\gamma \bar \Phi _{\alpha \alpha \beta
}\partial _\gamma b_\beta -12\partial _\alpha \bar \Phi _{\alpha
\beta \gamma }\partial _\gamma b_\beta  
\\\phan \non && 
{}-30\partial _\beta \Phi _{\alpha \alpha 0}\partial _\gamma \bar
\Phi _{\beta \gamma 0}+30\partial _\beta \Phi _{\alpha \alpha
0}\partial _\beta \bar \Phi _{\gamma \gamma 0}-30\partial _\beta \Phi
_{\alpha \alpha 0}\partial _\beta \bar \Phi _{000} \\\phan \non && 
{}-12\partial _\beta \Phi _{\alpha \alpha 0}\partial _\beta \bar
b_0{}+24\partial _\gamma \bar \Phi _{\beta \gamma 0}\partial _\alpha
\Phi _{\alpha \beta 0}-30\partial _\beta \bar \Phi _{\gamma \gamma
0}\partial _\alpha \Phi _{\alpha \beta 0}
 \\\phan \non &&
 {}+30\partial _\beta \bar \Phi _{000}\partial _\alpha \Phi
_{\alpha\beta 0} +\frac{42}5\partial _\alpha \Phi _{\alpha \beta
0}\partial _\beta \bar b_0{}-30\partial _\gamma \Phi _{\alpha \beta
0}\partial _\gamma \bar \Phi _{\alpha \beta 0} 
\\\phan\non && 
{}-12\partial _\beta \bar \Phi _{\alpha \alpha0}\partial _\beta b_0+
\frac{42}5\partial _\alpha \bar \Phi _{\alpha \beta 0}\partial _\beta
b_0-12\partial _\beta \bar \Phi _{\alpha 00}\partial _\beta b_\alpha  
\\\phan \non && 
{}-30\partial _\beta \Phi _{\alpha 00}\partial _\gamma \bar \Phi
_{\alpha \beta \gamma }+30\partial _\beta \Phi _{\alpha 00}\partial
_\beta \bar \Phi _{\alpha \gamma \gamma }-12\partial _\beta \Phi
_{\alpha 00}\partial _\beta \bar b_\alpha 
 \\\phan \non && 
{}-30\partial _\alpha \Phi _{\alpha 00}\partial _\gamma \bar \Phi
_{\beta \beta \gamma }{}+12\partial _\alpha \Phi _{\alpha 00}\partial
_\beta \bar b_\beta +12\partial _\alpha \bar \Phi _{\alpha
00}\partial _\beta b_\beta  
\\\phan \non && 
{}+20\partial _\alpha \bar \Phi _{000}\partial _\alpha \Phi
_{000}+12\partial _\alpha \bar \Phi _{000}\partial _\alpha
b_0{}+30\partial _\alpha \Phi _{000}\partial _\beta \bar \Phi
_{\alpha \beta 0} 
\\\phan \non && 
{}-30\partial _\alpha \Phi _{000}\partial _\alpha \bar \Phi _{\beta
\beta 0}+12\partial _\alpha \Phi _{000}\partial _\alpha \bar b_0-
\frac{36}5\partial _\beta b_\alpha \partial _\alpha \bar b_\beta
\\\phan \non && {}+
\frac{36}5\partial _\alpha b_\alpha \partial _\beta \bar b_\beta
+\frac{216} {25}\partial _\alpha b_0\partial _\alpha \bar
b_0+30\partial _\gamma h_{\alpha \beta }\partial _\gamma \bar
h_{\alpha \beta }-30\partial _\gamma h_{\alpha \beta }\bar \Phi
_{\alpha \beta \gamma }
 \\\phan \non &&{}
-30\partial _\gamma \bar h_{\alpha \beta }\Phi _{\alpha \beta \gamma
}{}+30\partial _\beta h_{\alpha \alpha }\partial _\gamma \bar
h_{\beta \gamma }-30\partial _\beta h_{\alpha \alpha }\partial _\beta
\bar h_{\gamma \gamma }+5\partial _\beta h_{\alpha \alpha }\partial
_\beta \bar \varphi
  \\\phan \non && 
{}+30\partial _\beta h_{\alpha \alpha }\partial _\beta \bar
h_{00}+15\partial _\beta h_{\alpha \alpha }\bar \Phi _{\beta
00}{}-15\partial _\beta h_{\alpha \alpha }\bar \Phi _{\beta \gamma
\gamma }-18\partial _\beta h_{\alpha \alpha }\bar b_\beta 
 \\\phan \non && 
{}-60\partial _\gamma \bar h_{\beta \gamma }\partial _\alpha
h_{\alpha \beta }+30\partial _\beta \bar h_{\gamma \gamma }\partial
_\alpha h_{\alpha \beta }-5\partial _\beta \bar \varphi \partial
_\alpha h_{\alpha \beta }{}-30\partial _\beta \bar h_{00}\partial
_\alpha h_{\alpha \beta } 
\\\phan \non && 
{}-60\partial _\alpha h_{\alpha \beta }\bar \Phi _{\beta
00}+60\partial _\alpha h_{\alpha \beta }\bar \Phi _{\beta \gamma
\gamma }+5\partial _\beta \bar h_{\alpha \alpha }\partial _\beta
\varphi +15\partial _\beta \bar h_{\alpha \alpha }\Phi _{\beta 00}
\\\phan \non && 
{}-15\partial _\beta \bar h_{\alpha \alpha }\Phi _{\beta \gamma
\gamma }-18\partial _\beta \bar h_{\alpha \alpha }b_\beta -5\partial
_\beta \varphi \partial _\alpha \bar h_{\alpha \beta }-60\partial
_\alpha \bar h_{\alpha \beta }\Phi _{\beta 00}
 \\\phan \non && 
{}+60\partial _\alpha \bar h_{\alpha \beta }\Phi _{\beta \gamma
\gamma }{}+ \frac{85}4\partial _\alpha h_{\alpha 0}\partial _\beta
\bar h_{\beta 0}+ \frac{325}8\partial _\alpha h_{\alpha 0}\bar \Phi
_{000}-\frac{255}8\partial _\alpha h_{\alpha 0}\bar \Phi _{\beta
\beta 0}
 \\\phan \non 
&&{}+\frac{63}4 \partial _\alpha h_{\alpha 0}\bar b_0-60\partial
_\beta h_{\alpha 0}\partial _\beta \bar h_{\alpha 0}{}+60\partial
_\beta h_{\alpha 0}\bar \Phi _{\alpha \beta 0}+60\partial _\beta \bar
h_{\alpha 0}\Phi _{\alpha \beta 0}
 \\\phan \non && {}+
\frac{325}8\partial _\alpha \bar h_{\alpha 0}\Phi
_{000}-\frac{255}8\partial _\alpha \bar h_{\alpha 0}\Phi _{\beta
\beta 0}+\frac{63}4\partial _\alpha \bar h_{\alpha 0}b_0-\frac
14\partial _\alpha \varphi \partial _\alpha \bar \varphi 
 \\\phan \non &&{}-5\partial _\alpha \varphi \partial _\alpha 
\bar h_{00}-\frac 52\partial _\alpha \varphi \bar \Phi _{\alpha
00}+\frac 52\partial _\alpha \varphi \bar \Phi _{\alpha \beta \beta
}-5\partial _\alpha \bar \varphi \partial _\alpha h_{00}-\frac
52\partial _\alpha \bar \varphi \Phi _{\alpha 00}
 \\\phan \non && 
{}+\frac 52\partial _\alpha \bar \varphi \Phi _{\alpha \beta \beta
}-45\partial _\alpha \bar h_{00}\Phi _{\alpha 00}+15\partial _\alpha
\bar h_{00}\Phi _{\alpha \beta \beta }+18\partial _\alpha \bar
h_{00}b_\alpha  
\\\phan \non && 
{}-30\partial _\alpha h_{00}\partial _\beta \bar h_{\alpha \beta
}{}+30\partial _\alpha h_{00}\partial _\alpha \bar h_{\beta \beta
}-45\partial _\alpha h_{00}\bar \Phi _{\alpha 00}+15\partial _\alpha
h_{00}\bar \Phi _{\alpha \beta \beta } \\\phan \non && 
{}+18\partial _\alpha h_{00}\bar b_\alpha +10\bar \Phi _{\alpha \beta
\gamma }\Phi _{\alpha \beta \gamma }{}+30\bar \Phi _{\beta 00}\Phi
_{\alpha \alpha \beta }- \frac{255}{16}\bar \Phi _{000}\Phi _{\alpha
\alpha 0}
 \\\phan \non &&{}+
\frac{165}{16}\bar \Phi _{000}\Phi _{000}+\frac{63}8\bar \Phi
_{000}b_0- \frac{255}{16}\bar \Phi _{\alpha \alpha 0}\Phi
_{000}{}-\frac{45}8\bar \Phi _{\alpha \alpha 0}b_0-\frac{45}8\Phi
_{\alpha \alpha 0}\bar b_0 
\\\phan  \non &&{}-30\bar
\Phi _{\beta \gamma \gamma }\Phi _{\alpha \alpha \beta }+
\frac{45}{16}\bar \Phi _{\beta \beta 0}\Phi _{\alpha \alpha 0}+30\bar
\Phi _{\alpha \beta \beta }\Phi _{\alpha 00}+\frac{63}8\Phi
_{000}\bar b_0 
\\\phan \non && 
{}+45\bar h_{00}h_{\alpha \alpha }-45\bar h_{00}h_{00}+45\bar
h_{\alpha \alpha }h_{00}-45\bar h_{\beta \beta }h_{\alpha \alpha
}+\frac{81}{20}\bar b_0b_0.
\end{eqnarray}

\section{}
\label{ap-ham}
Canonical Hamiltonian, corresponding to Lagrangian~\reff{l0s3b}, has
the form
\begin{eqnarray}
\label{ham3b}
{\cal H}&=&\frac 1{10}\bar p_{\alpha \beta \gamma }^\Phi p_{\alpha
\beta \gamma }^\Phi -\frac 3{50}\bar p_{\beta \gamma \gamma }^\Phi
p_{\alpha \alpha \beta }^\Phi +\frac 16\bar p_{000}^\Phi p_0^b+\frac
16\bar p_0^bp_{000}^\Phi +\frac 1{20}\bar p_{\alpha \alpha \beta
}^\Phi p_\beta ^b
\non\phan \\ && {}
+\frac 1{20}\bar p_\beta ^bp_{\alpha \alpha \beta }^\Phi {}+\frac
1{30}\bar p_{\alpha \beta }^hp_{\alpha \beta }^h-\frac 1{60}\bar
p_{\beta \beta }^hp_{\alpha \alpha }^h-\frac 5{36}\bar
p_0^bp_0^b+\frac 16\bar p_\alpha ^bp_\alpha ^b \phan
 \non\\ && {}
-\partial _\beta \Phi _{\alpha \alpha \beta }\bar p_{000}^\Phi  
-\partial _\beta \bar \Phi _{\alpha \alpha \beta }p_{000}^\Phi
{}-\partial _\alpha \Phi _{\alpha 00}\bar p_{000}^\Phi -\partial
_\alpha \bar \Phi _{\alpha 00}p_{000}^\Phi 
\phan \\ && {}
+\frac 35\partial _\alpha \Phi _{\alpha \beta0}\bar p_{\beta \gamma
\gamma }^\Phi  +\frac 35\partial _\alpha \bar \Phi _{\alpha \beta
0}p_{\beta \gamma \gamma }^\Phi +\frac 12\partial _\alpha h_{\alpha
0}\bar p_{\beta \beta }^h{}+\frac 12\partial _\alpha \bar h_{\alpha
0}p_{\beta \beta }^h 
\phan \non\\ && {}
+\frac 14\bar \Phi_{000}p_{\alpha \alpha }^h +\frac 14\Phi _{000}\bar
p_{\alpha \alpha }^h  +\bar \Phi _{\alpha \beta 0}p_{\alpha \beta
}^h+\Phi _{\alpha \beta 0}\bar p_{\alpha \beta }^h-\frac 34\bar \Phi
_{\alpha \alpha 0}p_{\beta \beta}^h{} 
\phan \non\\ && {}
-\frac 34\Phi _{\alpha \alpha 0}\bar p_{\beta \beta }^h
-\frac53\partial _\beta \Phi _{\alpha \alpha \beta }\bar p_0^b 
-\frac 53\partial _\beta \bar \Phi _{\alpha \alpha \beta }p_0^b-\frac
12\partial _\alpha \Phi _{\alpha \beta 0}\bar p_\beta ^b-\frac
12\partial _\alpha \bar \Phi _{\alpha \beta 0}p_\beta ^b{}
\phan \non\\ && {}
+\frac{10}3\partial _\alpha \Phi _{\alpha 00}\bar p_0^b 
+\frac{10}3\partial _\alpha \bar \Phi _{\alpha 00}p_0^b+\partial
_\alpha b_\alpha \bar p_0^b+\partial _\alpha \bar b_\alpha
p_0^b-\frac 12\partial _\alpha v_\alpha \bar p_0^b
\phan \non\\ && {}
-\frac 12\partial _\alpha \bar v_\alpha p_0^b -\frac 16\partial
_\alpha \bar v_0p_\alpha ^b-\frac 16\partial _\alpha v_0\bar p_\alpha
^b-2\partial _\alpha \bar \Phi _{000}\partial _\alpha v_0-2\partial
_\alpha \Phi _{000}\partial _\alpha \bar v_0 
\phan \non\\ && 
{}+2\partial _\beta \bar \Phi _{\alpha 00}\partial _\beta v_\alpha
{}+2\partial _\beta \Phi _{\alpha 00}\partial _\beta \bar v_\alpha
-2\partial _\beta \bar \Phi _{\alpha 00}\partial _\alpha v_\beta
-2\partial _\beta \Phi _{\alpha 00}\partial _\alpha \bar v_\beta  
\phan \non\\ && 
{}+6\partial _\alpha \Phi _{\alpha 00}\partial _\beta \bar v_\beta
{}+6\partial _\alpha \bar \Phi _{\alpha 00}\partial _\beta v_\beta
+2\partial _\beta \Phi _{\alpha \alpha 0}\partial _\beta \bar
v_0+2\partial _\beta \bar \Phi _{\alpha \alpha 0}\partial _\beta v_0 
\phan \non\\ && 
{}-4\partial _\alpha \Phi _{\alpha \beta 0}\partial _\beta \bar
v_0{}-4\partial _\alpha \bar \Phi _{\alpha \beta 0}\partial _\beta
v_0+2\partial _\alpha \Phi _{\alpha \beta \gamma }\partial _\gamma
\bar v_\beta +2\partial _\alpha \bar \Phi _{\alpha \beta \gamma
}\partial _\gamma v_\beta  
\phan \non\\ && 
{}-2\partial _\gamma \Phi _{\alpha \alpha \beta }\partial _\gamma
\bar v_\beta {}-2\partial _\gamma \bar \Phi _{\alpha \alpha \beta
}\partial _\gamma v_\beta +2\partial _\gamma \Phi _{\alpha \alpha
\beta }\partial _\beta \bar v_\gamma +2\partial _\gamma \bar \Phi
_{\alpha \alpha \beta }\partial _\beta v_\gamma  
\phan \non\\ && 
{}+20\partial _\alpha \bar \Phi _{000}\partial _\alpha \Phi
_{000}{}-30\partial _\beta \Phi _{\alpha \alpha 0}\partial _\beta
\bar \Phi _{000}-30\partial _\alpha \Phi _{000}\partial _\alpha \bar
\Phi _{\beta \beta 0} 
\phan \non\\ && 
{}+30\partial _\alpha \Phi _{000}\partial _\beta \bar \Phi _{\alpha
\beta 0}{}+30\partial _\beta \bar \Phi _{000}\partial _\alpha \Phi
_{\alpha \beta 0}-80\partial _\beta \bar \Phi _{\beta 00}\partial
_\alpha \Phi _{\alpha 00} 
\phan \non\\ && 
{}-30\partial _\beta \Phi _{\alpha \alpha 0}\partial _\gamma \bar
\Phi _{\beta \gamma 0}{}-30\partial _\beta \bar \Phi _{\gamma \gamma
0}\partial _\alpha \Phi _{\alpha \beta 0}+30\partial _\beta \Phi
_{\alpha \alpha 0}\partial _\beta \bar \Phi _{\gamma \gamma 0} 
\phan \non\\ && 
{}+24\partial _\gamma \bar \Phi _{\beta \gamma 0}\partial _\alpha
\Phi _{\alpha \beta 0}{}-30\partial _\gamma \Phi _{\alpha \beta
0}\partial _\gamma \bar \Phi _{\alpha \beta 0}+10\partial _\beta \Phi
_{\alpha \alpha \beta }\partial _\gamma \bar \Phi _{\gamma 00} 
\phan \non\\ && 
{}+10\partial _\alpha \Phi _{\alpha 00}\partial _\gamma \bar \Phi
_{\beta \beta \gamma }{}-30\partial _\alpha \Phi _{\alpha \beta
\gamma }\partial _\gamma \bar \Phi _{\beta 00}-30\partial _\beta \Phi
_{\alpha 00}\partial _\gamma \bar \Phi _{\alpha \beta \gamma }
\phan \non\\ && \phan
{}+30\partial _\gamma \bar \Phi _{\beta 00}\partial _\gamma \Phi
_{\alpha \alpha \beta }+30\partial _\beta \Phi _{\alpha 00}\partial
_\beta \bar \Phi _{\alpha \gamma \gamma }+10\partial _\delta \Phi
_{\alpha \beta \gamma }\partial _\delta \bar \Phi _{\alpha \beta
\gamma }  
\phan \non\\ && 
{}+10\partial _\beta \Phi _{\alpha \alpha \beta }\partial _\delta
\bar \Phi _{\gamma \gamma \delta }{}-30\partial _\alpha \Phi _{\alpha
\beta \gamma }\partial _\delta \bar \Phi _{\beta \gamma \delta
}+30\partial _\alpha \Phi _{\alpha \beta \gamma }\partial _\gamma
\bar \Phi _{\beta \delta \delta } 
\phan \non\\ && 
{}+30\partial _\delta \bar \Phi _{\beta \gamma \delta }\partial
_\gamma \Phi _{\alpha \alpha \beta }{}-30\partial _\gamma \bar \Phi
_{\beta \delta \delta }\partial _\gamma \Phi _{\alpha \alpha \beta
}+12\partial _\alpha \bar \Phi _{000}\partial _\alpha b_0 
\phan \non\\ && 
{}+12\partial _\alpha \Phi _{000}\partial _\alpha \bar b_0-12\partial
_\beta \bar \Phi _{\alpha 00}\partial _\beta b_\alpha {}-12\partial
_\beta \Phi _{\alpha 00}\partial _\beta \bar b_\alpha  
\phan \non\\ && 
{}-12\partial _\alpha \Phi _{\alpha 00}\partial _\beta \bar b_\beta
-12\partial _\alpha \bar \Phi _{\alpha 00}\partial _\beta b_\beta
{}-12\partial _\beta \Phi _{\alpha \alpha 0}\partial _\beta \bar b_0 
\phan \non\\ && 
{}-12\partial _\beta \bar \Phi _{\alpha \alpha 0}\partial _\beta
b_0+12\partial _\alpha \Phi _{\alpha \beta 0}\partial _\beta \bar
b_0+12\partial _\alpha \bar \Phi _{\alpha \beta 0}\partial _\beta b_0 
\phan \non\\ && 
{}-12\partial _\alpha \Phi _{\alpha \beta \gamma }\partial _\gamma
\bar b_\beta -12\partial _\alpha \bar \Phi _{\alpha \beta \gamma
}\partial _\gamma b_\beta +12\partial _\gamma \Phi _{\alpha \alpha
\beta }\partial _\gamma \bar b_\beta  
\phan \non\\ && 
{}+12\partial _\gamma \bar \Phi _{\alpha \alpha \beta }\partial
_\gamma b_\beta {}-30\partial _\alpha h_{\alpha \beta }\partial
_\beta \bar h_{00}-30\partial _\alpha h_{00}\partial _\beta \bar
h_{\alpha \beta } 
\phan \non\\ && 
{}+30\partial _\beta \bar h_{00}\partial _\beta h_{\alpha \alpha
}+30\partial _\alpha h_{00}\partial _\alpha \bar h_{\beta \beta
}-60\partial _\beta h_{\alpha 0}\partial _\beta \bar h_{\alpha 0} 
\phan \non\\ && 
{}+15\partial _\alpha h_{\alpha 0}\partial _\beta \bar h_{\beta
0}+30\partial _\gamma h_{\alpha \beta }\partial _\gamma \bar
h_{\alpha \beta }-60\partial _\alpha h_{\alpha \beta }\partial
_\gamma \bar h_{\beta \gamma }  
\phan \non\\ && 
{}+30\partial _\alpha h_{\alpha \beta }\partial _\beta \bar h_{\gamma
\gamma }+30\partial _\gamma \bar h_{\beta \gamma }\partial _\beta
h_{\alpha \alpha }-30\partial _\beta \bar h_{\gamma \gamma }\partial
_\beta h_{\alpha \alpha} 
\phan \non\\ && 
{}+\frac{75}2\partial _\alpha h_{\alpha 0}\bar \Phi
_{000}{}+\frac{75}2 \partial _\alpha \bar h_{\alpha 0}\Phi
_{000}-45\partial _\alpha h_{00}\bar \Phi _{\alpha 00}-45\partial
_\alpha \bar h_{00}\Phi _{\alpha 00}  
\phan \non\\ && \phan
{}+15\partial _\alpha h_{00}\bar \Phi _{\alpha \beta \beta
}{}+15\partial _\alpha \bar h_{00}\Phi _{\alpha \beta \beta
}+60\partial _\beta h_{\alpha 0}\bar \Phi _{\alpha \beta
0}+60\partial _\beta \bar h_{\alpha 0}\Phi _{\alpha
\beta 0} 
\phan \non\\ && 
{}-\frac{45}2\partial _\alpha h_{\alpha 0}\bar \Phi _{\beta \beta
0}{} -\frac{45}2\partial _\alpha \bar h_{\alpha 0}\Phi _{\beta \beta
0}-60 \partial _\alpha h_{\alpha \beta }\bar \Phi _{\beta
00}-60\partial _\alpha \bar h_{\alpha \beta }\Phi _{\beta 00} 
\phan \non\\ && \phan
{}+15\partial _\beta h_{\alpha \alpha }\bar \Phi
_{\beta 00}{}+15\partial _\beta \bar h_{\alpha \alpha }\Phi _{\beta
00}-30\partial _\gamma h_{\alpha \beta }\bar \Phi _{\alpha \beta
\gamma }-30\partial _\gamma \bar h_{\alpha \beta }\Phi _{\alpha \beta
\gamma } 
\phan \non\\ && 
{}+60\partial _\alpha h_{\alpha \beta }\bar \Phi _{\beta \gamma
\gamma }{}+60\partial _\alpha \bar h_{\alpha \beta }\Phi _{\beta
\gamma \gamma }-15\partial _\beta h_{\alpha \alpha }\bar \Phi _{\beta
\gamma \gamma }-15\partial _\beta \bar h_{\alpha \alpha }\Phi _{\beta
\gamma \gamma } 
\phan \non\\ && 
{}+\frac{35}4\bar \Phi _{000}\Phi _{000}{}-\frac{45}4\bar \Phi
_{000}\Phi _{\alpha \alpha 0}-\frac{45}4\bar \Phi _{\alpha \alpha
0}\Phi _{000}+30\bar \Phi _{\beta 00}\Phi _{\alpha \alpha \beta } 
\phan \non\\ && \phan
{}+30\bar \Phi _{\alpha \beta \beta }\Phi _{\alpha 00}{}-
\frac{45}4\bar \Phi _{\beta \beta 0}\Phi _{\alpha \alpha 0}+10\bar
\Phi _{\alpha \beta \gamma }\Phi _{\alpha \beta \gamma }-30\bar \Phi
_{\beta \gamma \gamma }\Phi _{\alpha \alpha \beta } 
\phan \non\\ && \phan\non
{} +\bar v_0\lambda _0+v_0\bar \lambda _0-\bar v_\alpha \lambda
_\alpha  {}-v_\alpha \bar \lambda _\alpha +\partial _\alpha \varphi
\bar \lambda _\alpha +\partial _\alpha \bar \varphi \lambda _\alpha\
. 
\end{eqnarray}

\end{document}